# History-Aware Trajectory $k$-Anonymization Using an FPGA-Based Hardware Accelerator for Real-Time Location Services

Hiroshi Nakano
*Graduate School of Science and Technology*
Keio University
Yokohama, Japan
nakano@west.sd.keio.ac.jp

Hiroaki Nishi
*Department of System Design Engineering, Faculty of Science and Technology*
Keio University
Yokohama, Japan
west@sd.keio.ac.jp

*Abstract*—Our previous work established the feasibility of FPGA-based real-time trajectory anonymization, a critical task for protecting user privacy in modern location-based services (LBS). However, that pioneering approach relied exclusively on shortest-path computations, which can fail to capture realistic travel behavior and thus reduce the utility of the anonymized data. To address this limitation, this paper introduces a novel, history-aware trajectory $k$-anonymization methodology and presents an advanced FPGA-based hardware architecture to implement it. Our proposed architecture uniquely integrates parallel history-based trajectory searches with conventional shortest-path finding, using a custom fixed-point counting module to accurately weigh contributions from historical data. This approach enables the system to prioritize behaviorally common routes over geometrically shorter but less-traveled paths. The FPGA implementation demonstrates that our new architecture achieves a real-time throughput of over 6,000 records/s, improves data retention by up to 1.2% compared to our previous shortest-path-only design, and preserves major arterial roads more effectively. These results signify a key advancement, enabling high-fidelity, history-aware anonymization that preserves both privacy and behavioral accuracy under the strict latency constraints of LBS.

## I. Introduction

The ubiquity of smartphones has resulted in the generation of vast amounts of user location data. Although this data is essential for applications such as traffic prediction, it also raises serious privacy concerns due to the potential for exposing sensitive personal information.

Trajectory anonymization—specifically, segment-based $k$-anonymization—addresses these concerns by publishing only road segments that are traversed by at least $k$ users, thereby preserving user privacy. However, software implementations struggle to meet the latency requirements of real-time location-based services (LBS), particularly as data volumes increase.

To address this challenge, hardware acceleration using FPGAs has gained attention. Our prior work [1], for example, introduced an FPGA accelerator for the segment-based $k$-anonymization method from [2], successfully achieving real-time throughput. However, existing accelerators rely on a fundamental simplification: the exclusive use of shortest-path computations. While computationally efficient, this approach often fails to reflect real-world travel behavior, as drivers may prefer major arterial roads over a geometrically shorter route. This discrepancy between assumed and actual paths can significantly reduce the utility of the anonymized data, leading to inaccurate traffic analyses.

This paper addresses this gap by introducing a novel, history-aware $k$-anonymization methodology designed to capture these realistic travel patterns. We present the first FPGA-based hardware architecture to implement this advanced approach in real time. Our design uniquely integrates parallel history-based trajectory searches with conventional shortest-path finding and uses a custom fixed-point counting module to accurately weigh contributions from historical data. This work enables the practical deployment of high-fidelity, history-aware anonymization in latency-sensitive LBS environments, preserving both privacy and behavioral accuracy under real-time constraints.

The main contributions of this paper are as follows:
1) A history-aware trajectory $k$-anonymization algorithm that increases data utility without weakening privacy.
2) An FPGA design featuring: (a) parallel shortest-path and history search, (b) Q16.16 fixed-point weighted counting, and (c) throughput exceeding 6,000 records/s.
3) An evaluation demonstrating that the proposed architecture meets real-time throughput requirements, improves data retention by up to 1.2% over a shortest-path-only FPGA, and operates with modest resource utilization.

## II. Related Work

### A. Trajectory Anonymization

A fundamental challenge in utilizing trajectory data lies in balancing data utility and user privacy. Various anonymization techniques have been proposed to address this issue, with $k$-anonymity serving as a widely recognized standard for trajectory data. However, the effectiveness, objectives, and associated trade-offs in terms of information loss and privacy risks differ significantly depending on the specific approach.

Location-based $k$-anonymization focuses primarily on protecting "stopping points" [3]–[6]. If fewer than $k$ users visit a given location, the corresponding data is generalized or suppressed. This often disrupts trajectory continuity and diminishes path detail, as non-stopping points may receive insufficient protection, and overgeneralization may reduce trajectory accuracy. A critical privacy concern is that adversaries with auxiliary information may infer individual movement patterns, particularly when overall trajectory shapes are not anonymized and thus remain uniquely identifiable.

Cluster-based $k$-anonymization groups similar trajectories into clusters of at least $k$ and releases a representative trajectory for each [7]–[10]. Although this method offers stronger privacy in theory, it often results in significant information loss, as individual trajectory details are abstracted away. This hampers fine-grained analyses such as detailed traffic flow estimation or behavioral pattern identification. Moreover, privacy risks persist if attackers can infer user attributes from clustering criteria or the choice of representative trajectories. Inaccurate cluster representation can also lead to analytical misinterpretations.

Another approach to privacy protection is differential privacy, which involves adding noise to the data. For example, the DP-Star [11] framework combines multiple techniques, such as representative point summarization and path correlation preservation, to achieve both differential privacy and high data utility. Furthermore, another proposed method uses deep neural networks to transform trajectory data into a latent space and then adds Laplace noise for anonymization [12]. However, while these differential privacy methods offer strong protection, the added noise degrades data utility for precise analyses like traffic flow estimation.

In contrast, segment-based $k$-anonymization [1], [2]—which we adopt in this study—offers a superior balance. This method maps location data onto road segments and publishes only those traversed by at least $k$ users. By doing so, it preserves detailed movement information at a finer granularity than location- or cluster-based approaches, facilitating analyses such as road usage frequency. Although segments traversed by fewer than $k$ users are suppressed, this intentional information loss enhances privacy by preventing the exposure of unique travel patterns and making full trajectory reconstruction more difficult for attackers. Overall, segment-based anonymization strikes a practical balance between data utility and privacy protection by addressing the limitations of alternative methods.

*B. Hardware Acceleration for Trajectory Anonymization*

The computational demands of processing large-scale trajectory data, especially under the stringent latency requirements of real-time applications, have spurred significant interest in hardware acceleration.

FPGAs are particularly well-suited to this domain due to their reconfigurability and support for fine-grained parallelism.

As a notable example, our earlier study [1] proposed an FPGA-based hardware accelerator for the segment-based $k$-anonymization method introduced by [2]. This architecture incorporated dedicated modules for key anonymization stages, including node approximation (utilizing hash tables for efficient map access), trajectory search (implementing Dijkstra's algorithm [13]), segment generation, and segment counting. This work demonstrated that real-time throughput for segment-based $k$-anonymization is achievable.

However, such prior efforts in hardware acceleration have focused exclusively on optimizing anonymization via shortest-path computations. The challenge and opportunity of integrating user movement history into a hardware-accelerated $k$-anonymization pipeline—to improve the realism and utility of anonymized data while still meeting real-time constraints—remains unexplored. This work addresses this gap by proposing a novel, history-aware anonymization methodology and a corresponding FPGA architecture tailored for its efficient execution.

III. PROPOSED METHODOLOGY

This section details our novel trajectory $k$-anonymization methodology that integrates user movement history to enhance data utility while ensuring privacy protection and real-time performance.

Key terms used in this paper are defined as follows: a **Node** represents an intersection on a digital map; an **Edge** is a road segment connecting two nodes; a **Path** is a sequence of nodes and edges, typically without temporal information; a **Trajectory** denotes movement as a time-ordered sequence of spatial points or, in this context, a sequence of nodes representing user travel; and a **Segment** is the minimal unit of a trajectory, consisting of two neighboring nodes.

The proposed system processes user location data through a pipeline designed to incorporate movement history into $k$-anonymization, ultimately outputting privacy-preserving trajectory data. The main processing steps are as follows:

*A. Node Approximation*

The system first receives raw location data, comprising user IDs and spatio-temporal coordinates (latitude, longitude, and timestamp). Each location point is mapped to its nearest representative node on a digital map. This process transforms continuous movement data into discrete node sequences, which form the basis for subsequent trajectory processing.

*B. Trajectory Estimation*

For each pair of approximated start and end nodes derived from the previous step, the system estimates the user's most likely travel paths. This estimation considers two sources of information:

*1) Shortest Path Search:* The geographically shortest path between the start and end nodes is computed using a standard pathfinding algorithm (Dijkstra's method [13]). The resulting node sequence serves as a baseline candidate trajectory.

*2) History-Based Trajectory Search:* This phase queries a pre-constructed historical trajectory database to retrieve past travel patterns between the same start and end nodes. This database is crucial for integrating user behavioral patterns into the anonymization.

**Database Construction and Rationale:** Real-world location data is often sparse or irregularly sampled. To address this, our offline database construction process uses raw historical location data to infer plausible routes. For each user, the system processes their time-ordered sequence of sparse location points and applies a shortest-path algorithm to fill the gaps between consecutive points. This process transforms a user's sparse location samples into a complete, connected trajectory.

It is critical to understand how this database, built using shortest-path inference, can yield non-shortest-path trajectories for the final anonymization. While the path between any two consecutive sparse points is the shortest, the full trajectory is the concatenation of these segments, following the user's entire sequence of recorded locations. For example, if a user's sparse data points follow a major highway, the concatenated path will reflect this preference, even if a geometrically shorter side-road path exists between the overall start and end points of the trip. This allows the database to capture realistic, behavior-driven routes that are not necessarily the single, globally shortest path.

**History Search and Filtering:** During the online history search, the current start/end node pair is used to query this database. The system retrieves all historical trajectories from any user that match the given start and end nodes. Optionally, a hop count filter can be applied to exclude implausibly long historical routes. The system first determines the hop count of the computed shortest path to serve as a baseline. Historical trajectories whose hop count exceeds this baseline by more than a user-configurable limit $\Delta_h$ are discarded. This feature allows for tuning the plausibility of accepted routes; in our experiments, we use $\Delta_h = 5$, as it was empirically found to provide a good balance, but this can be adjusted for different use cases.

## C. Candidate Trajectory Determination

Based on the outputs from the trajectory estimation phase, the system decides which trajectory information to use for anonymization. If one or more valid historical trajectories are retrieved, this set is selected for further processing. If no suitable historical trajectories are found, the system defaults to using the shortest path. This approach prioritizes realistic user movement patterns derived from historical data when available.

## D. Anonymization with History-Aware Segment Counting

The trajectory information determined in the previous step is used to update segment traversal counts for $k$-anonymization. The selected trajectory (or set of trajectories) is decomposed into its constituent segments. For each segment, the system updates a count of distinct users estimated to have traversed it. The method of updating depends on whether historical trajectories or a shortest path was selected:

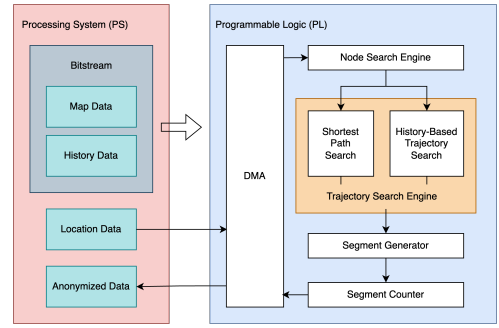

Fig. 1. Overall architecture of the proposed history-aware anonymization accelerator. The Processing System (PS) manages data flow, while the Programmable Logic (PL) performs the core tasks: node searching, parallel trajectory estimation (shortest-path and history-based), and anonymized segment counting.

**If a set of historical trajectories is used:** The contribution of the current user is distributed among all valid historical paths found. Let $h$ be the number of valid historical trajectories retrieved from the database. Each segment in any of these $h$ trajectories has its traversal count incremented by a weight of $1/h$.

**If the shortest path is used:** When the shortest path is used, each of its segments has its traversal count incremented by 1.

Finally, only those segments whose accumulated (and potentially weighted) user counts meet or exceed the threshold $k$ are published as anonymized data. Segments with counts below $k$ are suppressed to ensure $k$-anonymity.

## IV. IMPLEMENTATION

This section details the FPGA implementation of the proposed history-aware $k$-anonymization system. Figure 1 illustrates the overall system architecture. The Processing System (PS) manages data flow and control, communicating with the Programmable Logic (PL) where the core anonymization tasks are accelerated. Map and historical trajectory data are preloaded, and input location data is transferred via Direct Memory Access (DMA). The PL integrates a Node Search Engine, a Trajectory Search Engine (performing both shortest-path and history-based searches), segment generation, and history-aware segment counting modules.

### A. Trajectory Search Engine

The trajectory estimation process described in Section III-B is realized in hardware by the Trajectory Search Engine. Given a pair of location points (approximated as start and end nodes), the Trajectory Search Engine identifies candidate trajectories. This involves the parallel execution of shortest-path computation and history-based trajectory retrieval.

*1) Shortest Path Search:* This module computes the geographically shortest path between the input start and end nodes using Dijkstra's algorithm. It outputs the sequence of node IDs constituting the shortest path and its total hop count. This path serves as a fallback when no suitable historical trajectories are available.

*2) History-Based Trajectory Search:* This module leverages past movement records, stored in on-chip Block RAM (BRAM), to identify realistic and frequently used travel patterns that may deviate from the shortest path. To maximize hardware efficiency, the historical trajectory database $H$ is structured as a single, large, time-ordered log of $(n, u)$ tuples, where $n$ is a node ID and $u$ is a user ID.

The search process, detailed in Algorithm 1, is initiated with a start node $n_s$, an end node $n_e$, and a maximum allowable hop count $maxHop$ derived from the shortest-path search. The core of the search logic is a single-pass process that scans the entire history database $H$ from beginning to end. Upon finding an entry that matches $n_s$, dedicated hardware logic transitions to a "path tracking" state. In this state, it latches the current user ID and sequentially tracks subsequent node visits by the same user to construct a temporary path.

The tracking process continues until one of the following termination conditions is met: (1) the user ID changes, which is interpreted as the end of one user's movement sequence; (2) the hop count of the path under construction exceeds $maxHop$, which prunes implausibly long routes; or (3) the path loops back to the start node $n_s$, which avoids meaningless cycles. If the end node $n_e$ is reached within these constraints during tracking, the path is deemed a valid historical trajectory. Its complete node sequence is stored as a result, and a hit counter $h$ is incremented.

This full-scan approach, while computationally expensive in software, is extremely efficient in hardware. It maps directly to a simple state machine and a single linear access to BRAM, maximizing data locality and eliminating the random memory access that FPGAs handle poorly. As a result, the search completes in a deterministic number of clock cycles that depends only on the size of the history database, guaranteeing the predictable and stable performance essential for real-time systems.

As detailed in our methodology (Section III), it is worth reiterating that while our database construction uses shortest-path interpolation, the resulting historical trajectories accurately capture real user movement. This is because the process connects the observed sequence of location points for each user, ensuring the final generated path aligns with their actual, behavior-driven route, not merely a geometric shortest path.

*3) Trajectory Selection:* After the parallel searches complete, the engine selects which paths to forward. The history search, which requires a full scan of its BRAM, dictates the overall search time. Once it finishes, the engine inspects the hit count $h$. If $h > 0$, the set of $h$ historical paths is used for anonymization. If $h = 0$, the system defaults to the pre-computed shortest path. This logic ensures that historical data is prioritized, aligning with the methodology in Section III-C.

### B. Anonymization Module

A key challenge in implementing our history-aware methodology is handling the weighted segment counts. Unlike

---

**Algorithm 1** History-Based Trajectory Search

**Require:** $n_s, n_e, \Delta_h, H$ {$n_s$: start, $n_e$: end, $\Delta_h$: hop limit, $H$: history DB}
**Ensure:** $P, h$ {$P$: paths, $h$: hit count}
1: $sp \leftarrow \text{DijkstraHopCount}(n_s, n_e)$
2: $maxHop \leftarrow sp + \Delta_h$
3: $P \leftarrow [\,], h \leftarrow 0$
4: **for** $i = 1$ **to** $|H|$ **do**
5:   **if** $H[i].n = n_s$ **then**
6:     $current\_user \leftarrow H[i].u$
7:     $cur \leftarrow [n_s], c \leftarrow 0$
8:     **for** $j = i + 1$ **to** $|H|$ **do**
9:       **if** $H[j].u \neq current\_user$ **or** $H[j].n = n_s$ **or** $c \geq maxHop$ **then**
10:         **break**
11:       **end if**
12:       $c \leftarrow c + 1$
13:       $cur.\text{append}(H[j].n)$
14:       **if** $H[j].n = n_e$ **then**
15:         $P.\text{append}(cur), h \leftarrow h + 1$
16:         **break**
17:       **end if**
18:     **end for**
19:   **end if**
20: **end for**
21: **return** $P, h$ =0

---

shortest-path-only approaches where counts are simple integers, our method requires accumulating fractional weights $(1/h)$. To address this efficiently in hardware, the Anonymization Module employs fixed-point arithmetic. The module receives either a set of historical trajectories or a single shortest path, decomposes them into segments, and updates traversal counts. To accumulate these potentially fractional values, we designed a custom **Segment Counter** that utilizes BRAM and 32-bit fixed-point arithmetic in Q16.16 format. Each segment $(a, b)$ is hashed to a BRAM address where its traversal count is stored.

## V. EVALUATION

This section evaluates the performance of the proposed history-aware trajectory $k$-anonymization methodology and its hardware accelerator, hereafter referred to as the **Proposed Architecture**. We focus on three key aspects: anonymization quality (data utility), processing performance, and hardware resource usage. For comparison, we use a previously developed shortest-path-only hardware accelerator [1], referred to as the **Baseline Architecture**.

### A. Experimental Environment and Datasets

*1) Hardware Platform:* The Proposed and Baseline Architectures were implemented on the M-KUBOS FPGA platform [14], which is equipped with a Xilinx Zynq UltraScale+ MPSoC (XCZU19EG-FFVC1760-2-I). The Processing System (PS) communicates with the Programmable Logic (PL) using the PYNQ framework.

*2) Map and Location Data:* The map data covers a 4 km² area around Kita-Urawa Station in Saitama, Japan, obtained

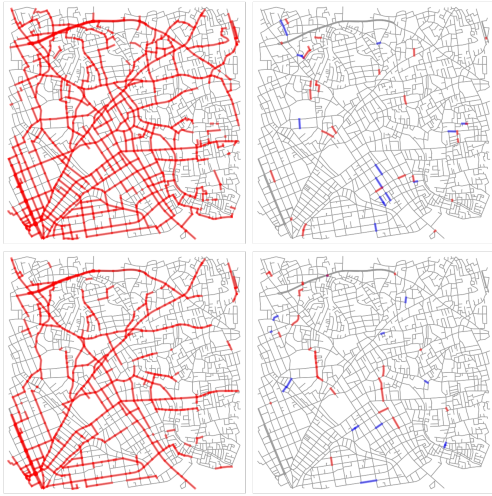

Fig. 2. Anonymization results for $k = 16$ (top) and $k = 32$ (bottom). Left panels: Road network published by the history-aware method. Right panels: Differential analysis, where red paths are unique to the history-aware method and blue paths are unique to the shortest-path-only method.

from OpenStreetMap (OSM) via OSMnx. This dataset comprises 4,500 nodes and 5,100 edges. Anonymization was performed on location data from Agoop Corp., using records from 10:00-11:00 AM (51,103 data points) as the primary input. The historical trajectory database was constructed from location data from the preceding hour (9:00-10:00 AM, 49,083 data points).

### B. Anonymization Quality and Data Utility

*1) Visual Analysis of Anonymization Results:* To highlight the qualitative difference between the two approaches, Figure 2 presents the anonymization results. The figure compares the published road networks for $k = 16$ (top) and $k = 32$ (bottom). The left panels show the output of our Proposed Architecture. The right panels provide a differential view: paths in red were published *only* by our history-aware method, while paths in blue were published *only* by the Baseline (shortest-path) Architecture.

The results clearly show that the paths unique to the history-aware method (red) consistently correspond to major arterial roads and common routes. In contrast, paths unique to the shortest-path method (blue) tend to be minor residential streets or geometrically shorter but less-traveled shortcuts. This visually confirms that our history-aware approach successfully filters out unrealistic shortcuts and better preserves the core traffic corridors essential for high-utility traffic analysis, providing a more accurate representation of real-world movement.

*2) Data Retention Rate:* The data retention rate, defined as the percentage of published segments after anonymization relative to the total number of unique segments in the input, serves as a metric for information loss. Figure 3 shows the retention rate as a function of $k$.

The Proposed Architecture demonstrates a slightly higher data retention rate, improving by up to 1.2% over the Baseline.

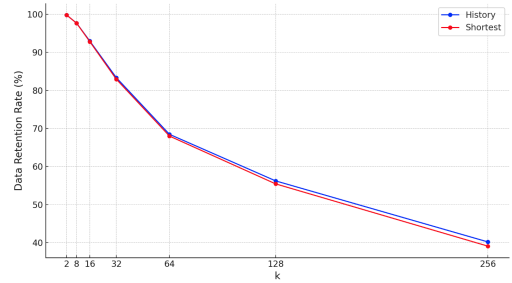

Fig. 3. Data retention rate comparison for varying values of $k$. The Proposed Architecture maintains a higher retention rate than the Baseline, and the advantage grows as privacy requirements ($k$) become stricter.

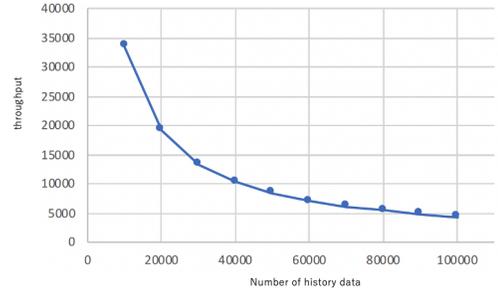

Fig. 4. Throughput vs. Historical Dataset Size. The system maintains the real-time target of 6,000 records/s, achieving this benchmark with datasets as large as 70,000 entries.

While numerically modest, this improvement is significant when viewed alongside the qualitative results. The history-aware method consolidates traffic counts onto behaviorally common routes. By focusing trajectories on popular, real-world paths, it ensures that these key segments are more likely to surpass the anonymity threshold, especially at higher $k$ values where privacy requirements are stricter. This prevents the dispersal of counts across many geometrically short but less-frequented segments, which are then suppressed, leading to a more coherent and useful anonymized road network.

### C. Processing Performance

*1) Throughput Analysis:* Throughput, measured in location records processed per second, is a key metric for real-time services. To process the location data from a region with the population density of Saitama City (6,089 people/km$^2$) within one minute, a target throughput of approximately 6,000 records/s is necessary. As shown in Figure 4, the throughput of the Proposed Architecture is inversely proportional to the size of the history database, as the history search time scales linearly with its size. The architecture is engineered to sustain a throughput of at least 6,000 records/s, a benchmark it successfully meets for historical datasets up to 70,000 trajectories, confirming its suitability for real-time deployment.

*2) Performance Overhead vs. Baseline Architecture:* While providing higher data utility, the history-aware logic introduces a performance overhead. The processing time of the Proposed Architecture is approximately 3.33 times longer than that of the Baseline, independent of the input data size. This overhead is an expected trade-off for the enhanced data utility, and the resulting throughput remains well within the requirements for real-time deployment.

### D. Impact of Hop Count Limit Feature

We evaluated the impact of the hop count limit feature, which is designed to improve result plausibility by excluding excessively long historical paths. Our analysis shows that applying this filter has a nuanced effect. At lower $k$ values (e.g., $k < 32$), the filter prunes a noticeable number of outlier segments, reducing the total count of published segments by up to 3%. This demonstrates its effectiveness in removing less common, circuitous routes. However, as $k$ increases, the impact diminishes rapidly, becoming negligible for $k \geq 64$. This occurs because at higher anonymity levels, the results are already dominated by major, frequently traversed paths that would not be filtered. This feature thus provides a valuable tuning parameter for refining data utility, particularly for use cases focused on lower $k$ values.

### E. Resource Utilization and Clock Frequency

Table I compares the FPGA resource utilization of the Proposed and Baseline Architectures. The Proposed Architecture requires more resources, particularly BRAM, due to the need to store the historical trajectory database and its associated search logic. The Segment Counter also uses more resources due to its wider BRAM and the logic for handling fixed-point arithmetic.

TABLE I
FPGA RESOURCE UTILIZATION COMPARISON

| Resource | Baseline [1] | Proposed | Difference |
|---|---|---|---|
| LUTs | 11.49% | 12.61% | +1.12% |
| FFs | 9.24% | 10.60% | +1.36% |
| BRAMs | 27.03% | 39.74% | +12.71% |

Despite the increased size, the design was successfully placed and routed, achieving a maximum operating frequency of 107 MHz. This confirms that the history-aware methodology can be implemented efficiently on modern FPGAs, with ample resources remaining for further expansion or integration of additional features.

### VI. CONCLUSION

This paper introduced a history-aware trajectory $k$-anonymization methodology and its real-time FPGA accelerator to address the data utility limitations of conventional shortest-path-only approaches. Our architecture integrates a parallel search engine for historical routes alongside shortest-path finding and uses fixed-point arithmetic for weighted segment counting, enabling it to generate more realistic anonymized data. The evaluation confirmed that the proposed architecture achieves real-time throughput of over 6,000 records/s while improving data utility, as demonstrated by up to a 1.2% higher data retention rate and better preservation of major arterial roads. The design's modest resource utilization confirms its practicality. This work successfully demonstrates that integrating movement history into a hardware-accelerated pipeline enables a new class of anonymization systems that provide a superior balance between user privacy and data fidelity for real-time LBS.

ACKNOWLEDGMENT

This work was supported by the JST SIP (Grand Number JPJ012207). The authors also express their gratitude to the MAFF commissioned project (Grant Number JPJ009819).